\documentclass[conference]{IEEEtran}

\usepackage{stfloats}
\usepackage{pdfsync}
\usepackage{cite}
\usepackage{amsthm, amssymb}
\usepackage[cmex10]{amsmath}

\usepackage{graphicx}   
\usepackage{tikz}					
\usepackage{pgflibraryplotmarks}		
\usepackage{pgflibraryarrows}

\graphicspath{{Figures/}}
\DeclareGraphicsExtensions{.pdf}	

\IEEEoverridecommandlockouts

\ifCLASSINFOpdf
\else
\fi

\begin{document}

\title{An Analog Baseband Approach for Designing Full-Duplex Radios}

\author{\IEEEauthorblockN{Brett Kaufman\IEEEauthorrefmark{1}\IEEEauthorrefmark{2}, Jorma Lilleberg\IEEEauthorrefmark{1}\IEEEauthorrefmark{2}\IEEEauthorrefmark{3},  and Behnaam Aazhang\IEEEauthorrefmark{1}\IEEEauthorrefmark{2}}
\IEEEauthorblockA{\IEEEauthorrefmark{1}Center for Multimedia Communication, Rice University, Houston, Texas, USA
\\
\IEEEauthorblockA{\IEEEauthorrefmark{2}Centre for Wireless Communications, University of Oulu, Oulu, Finland
\\
\IEEEauthorblockA{\IEEEauthorrefmark{3}Renesas Mobile, Oulu, Finland}}}
\thanks{This work is funded in part by a NSF grant, by Renesas through a research contract, and the Academy of Finland through the Co-Op grant.}
}
  
\maketitle  

\begin{abstract}
Recent wireless testbed implementations have proven that full-duplex communication is in fact possible and can outperform half-duplex systems.  Many of these implementations modify existing half-duplex systems to operate in full-duplex.  To realize the full potential of full-duplex, radios need to be designed with self-interference in mind.  In our work, we use an experimental setup with a patch antenna prototype to characterize the self-interference channel between two radios.  In doing so, we form an analytical model to design analog baseband cancellation techniques.  We show that our cancellation scheme can provide up to 10 dB improved signal strength, 2.5 bps/Hz increase in rate, and a $10^4$ improvement in BER as compared to the RF only cancellation provided by the patch antenna.  
\end{abstract}



%

\section{Introduction}   
\begin{figure}[htp]
\begin{center} 
  \includegraphics[width=0.4\textwidth]{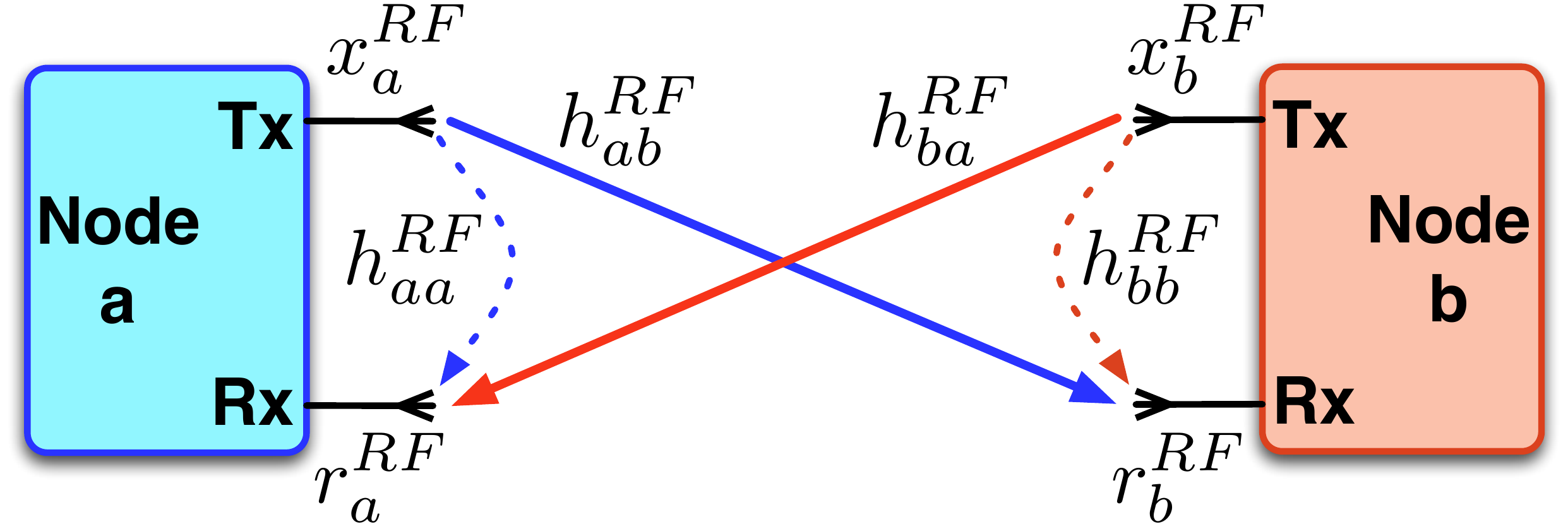}
  \vspace{-12pt}
  \caption[fig:chan_mag]{Two-user bidirectional full-duplex link showing self-interference channels with dashed arrows and data channels with solid arrows.} 
  \label{fig:network}
\end{center} 
\end{figure}

Wireless full-duplex communication in which a terminal can simultaneously transmit and receive in the same frequency band was first demonstrated in radar systems \cite{FD_Radar} as early as the 1940's.  Then in the 1980's, cellular networks utilized full-duplex in repeaters \cite{FD_Repeater} to extend cellular coverage.  Not until recently in 2010 was a bidirectional point-to-point full-duplex link, shown in Fig.~\ref{fig:network}, demonstrated with experimental testbeds \cite{2010_Melissa_Asilomar}, \cite{2010_Stanford_Mobicom}.  However, insufficient levels of self-interference cancellation prevented the expected doubling of spectral efficiency as compared to half-duplex communications from being achieved.  

Current self-interference cancellation techniques can be classified into two main techniques: \emph{Passive Suppression} and \emph{Active Cancellation}.  Passive techniques attempt to increase the isolation between the transmit and receive antennas and are agnostic to the signal characteristics of the self-interference signal.  A combination of directional isolation, absorptive shielding, and cross-polarization in the transmit and receive antennas was used in \cite{2013_Evan_TWC_passive}.  Another experimental setup \cite{2012_MIDU} used multiple transmit antennas to create a null point at the receive antenna.  A novel antenna design in \cite{2010_Patch_Sweden} isolates the transmit and receive streams with two dual-polarized patch antennas.  

Active techniques enable a terminal to use the knowledge of it's own self-interference signal to generate a cancellation signal that can can be subtracted from the received signal.  An experimental setup using the WARP platform \cite{2012_Melissa_TWC} used an extra transmit chain to generate an up-converted RF cancellation signal that was then subtracted from the incoming signal at the receive antenna.  A recent work in \cite{2013_Stanford_Sigcomm} proposes active circuitry that samples the RF self-interference signal and uses sinc interpolation to generate the cancellation signal. 

Instead of an active or passive labeling for the self-interference cancellation technique, we can classify the cancellation technique based on which point along the transceiver chain does the cancellation occur.  All of the above listed techniques are implemented in the analog RF stage and are primarily done so in order to not saturate the low noise amplifier.  An additional reason for focusing on the analog RF stage of the transceiver is the relative ease in which circuit modifications and additions can be connected to the existing radio design.  Work in \cite{2012_Melissa_TWC} and \cite{2013_Stanford_Sigcomm} demonstrate active circuitry that can connect to the transmitter and receiver chains respectively.  

It is due to the above two reasons why the \emph{analog baseband} stage of the transceiver has been largely ignored for self-interference cancellation.  Providing an additional stage of analog cancellation just before the analog-to-digital converter (ADC) would increase the dynamic range of the ADC and thus provide better resolution of the desired signal over the self-interference signal.  The work in this paper will show that adding a complementary analog baseband self-interference cancellation stage to analog RF self-interference cancellation can significantly improve the total cancellation achieved and help close the gap between the experimental implementations and the theoretical expectations.  

The remainder of this paper is organized as follows.  In Section~\ref{sec:system_model} we define the transceiver model and signal model.  In Section~\ref{sec:self_int}, we characterize the self-interference channel.  Then in Section~\ref{sec:cancel} we provide details of the proposed cancellation scheme and quantify its performance.  We then evaluate a two-terminal full-duplex link in Section~\ref{sec:FDlink} and then finish with concluding remarks in Section~\ref{sec:conclusion}.

\section{System Model}
\label{sec:system_model}
\begin{figure}[htp]
\begin{center} 
  \includegraphics[width=0.4\textwidth]{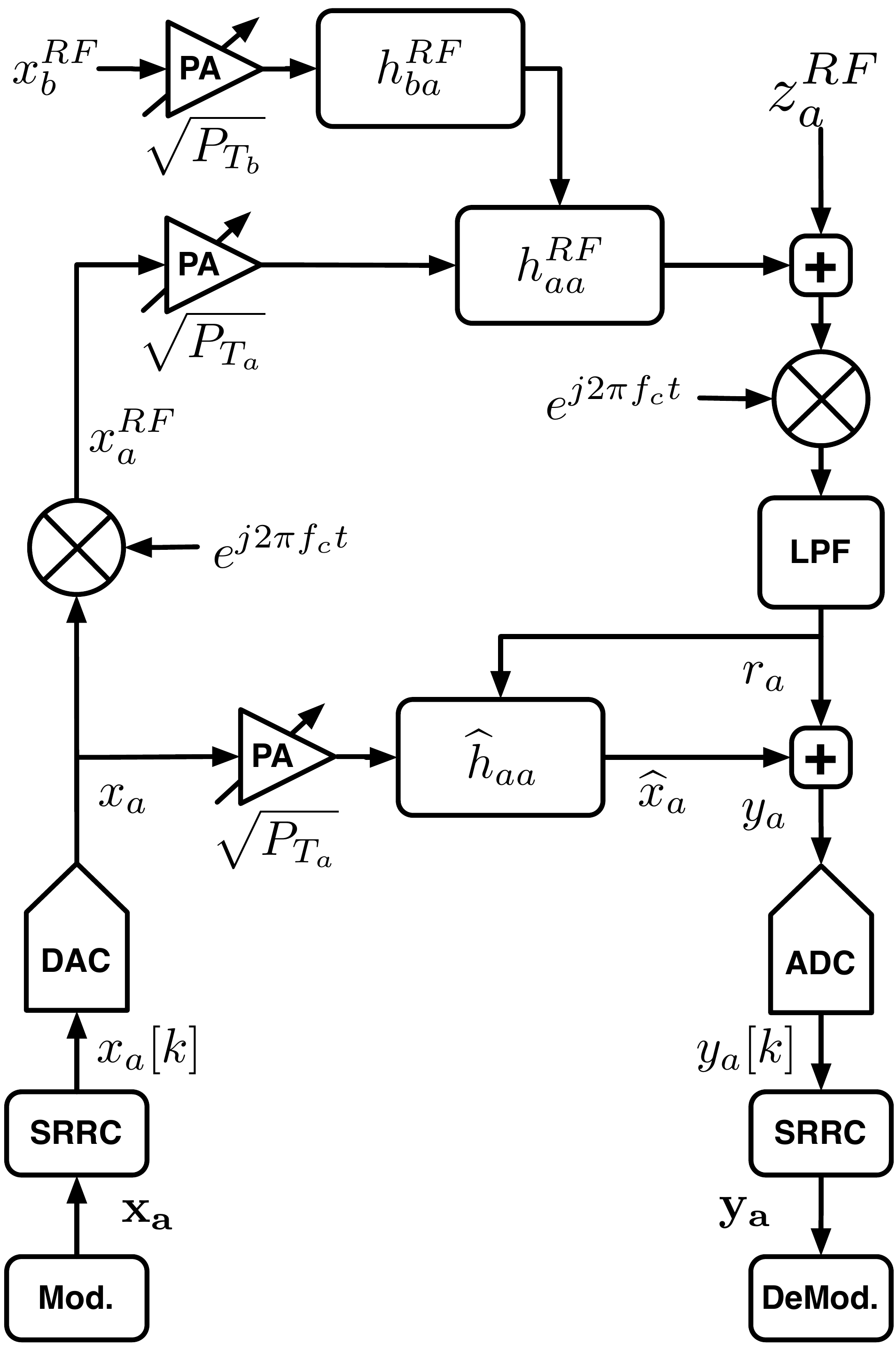}
  \vspace{-8pt}
  \caption[fig:system]{Functional block diagram of a full-duplex transceiver from the perspective of terminal $a$. } 
  \label{fig:system}
\end{center} 
\end{figure}

We consider the two-terminal point-to-point full-duplex link shown in Fig.~\ref{fig:network} where terminals $a$ and $b$ are communicating with each other using the same temporal and frequency resources.  Each terminal has a single transmit and receive antenna.  We now refer to the functional block diagram in Fig.~\ref{fig:system} as we derive the signal model.  We note that the block diagram is from the perspective of terminal $a$ and that everything is identical for terminal $b$.  

At the transmitter side, the baseband signals of bandwidth $BW$ are first modulated (Mod.) into $M$-PSK symbols $\mathbf{x_a}$.  We assume average unit energy symbols with $E[||\mathbf{x_{a}}||^2] = 1$ where $E[\cdot ]$ is used to denote the statistical expectation.  The symbols are then pulse-shaped using a square-root-raised-cosine (SRRC) filter and the output digital samples $x_a[k]$ serve as input to the Digital-to-Analog Converter (DAC).  We assume ideal DACs such that the output baseband time domain signal $x_a(t)$ satisfies $x_a[k] \cong x_a(t)$.  As our focus will be on the analog domain, we remove the time notation $t$ for simplification and simply refer to $x_a(t)$ as $x_a$.  We will maintain this assumption for all other time domain signals.  The analog baseband signal $x_a$ is then up-converted to the carrier frequency $f_c$ yielding $x_a^{RF} = x_a e^{j 2\pi f_c t}$.  The signal is then amplified with signal power $P_{T_a}$ by a power amplifier (PA).

At the receiver side, after down-conversion from $f_c$ and low-pass filtering (LPF), the received baseband time domain signal $r_a$ can be expressed as
\begin{equation}
r_a = \sqrt{P_{T_b}}h_{ba}x_b + \sqrt{P_{T_a}}h_{aa}x_{a} + z_a,
\label{eq:r_bb}
\end{equation}
where $x_b$ is the signal-of-interest transmitted over the wireless channel $h_{ba}$ and $x_{a}$ is the self-interference signal transmitted over the self-interference channel $h_{aa}$.  The received signal is corrupted by additive white Gaussian noise $z_a \sim \mathcal{CN}(0,\sigma_z^2)$.  We note that the received analog baseband signal in (\ref{eq:r_bb}) is the analytical equivalent for the received passband signal at carrier frequency $f_c$.  

Just after down-conversion and low-pass filtering, an estimate of the self-interference signal $\widehat{x_a}$ is added to the received baseband signal giving 
\begin{equation}
y_a = r_a + \widehat{x}_a,
\label{eq:y_bb}
\end{equation}
which in turn serves as input to the ADC.  Output digital samples $y_a[k]$ from the ADC pass through a receiver side SRRC yielding symbols $\mathbf{y_a}$ which can be finally demodulated (DeMod).  

\section{Self-Interference Model}
\label{sec:self_int}
We now provide details about the RF self-interference channel $h_{aa}$.  We utilize a four-layer patch antenna prototype designed by \cite{RISC_WEB} and similar in design to \cite{2011_patch_antenna}.  The patch antenna isolates the transmit and receive antennas from each other in a single form factor.  Because the isolation is not perfect, an attenuated version of the self-interference signal from terminal $a$, the coupling signal, passes thru the antenna from the transmitter side to the receiver side and mixes with the incoming desired signal from terminal $b$.    

\begin{figure}[htp]
\begin{center} 
  \includegraphics[width=0.4\textwidth]{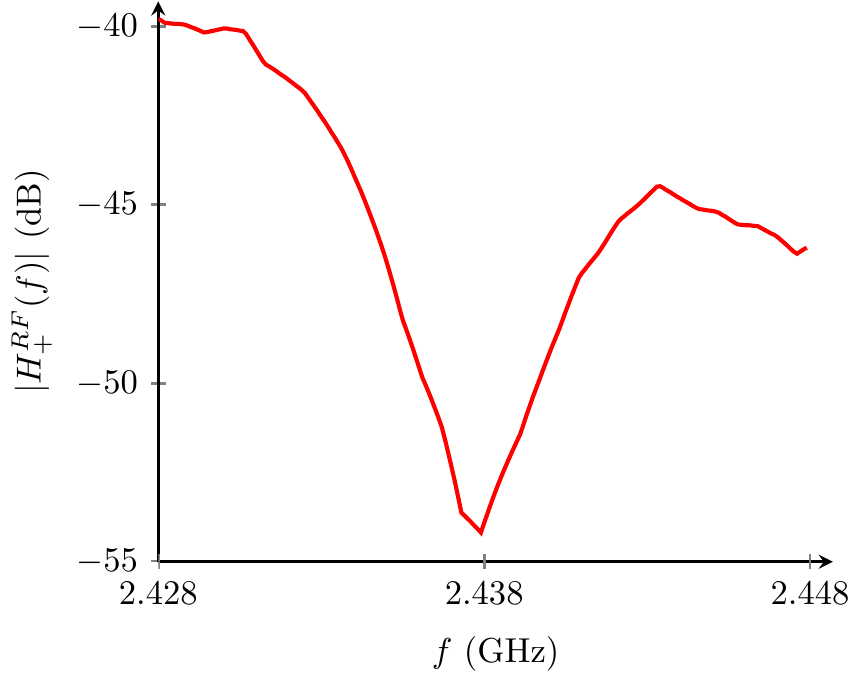}
  \vspace{-8pt}
  \caption[fig:chan_mag]{Isolation measurements for the full-duplex patch antenna prototype.} 
  \label{fig:chan_mag}
\end{center} 
\end{figure}
\begin{figure}[htp]
\begin{center} 
  \includegraphics[width = 0.4\textwidth]{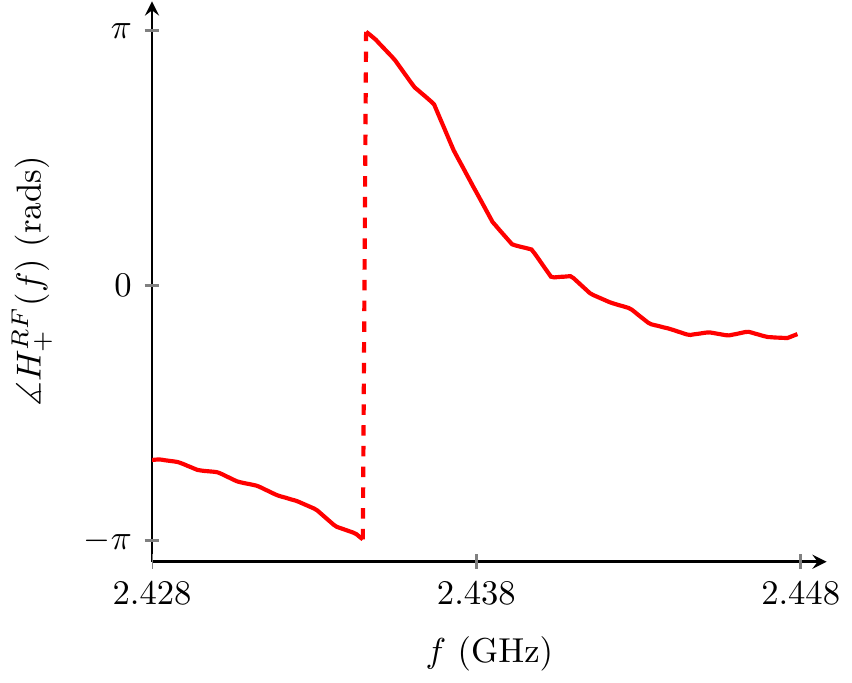}
  \vspace{-8pt}
  \caption[fig:chan_ang]{Phase measurements for the full-duplex patch antenna prototype.} 
  \label{fig:chan_ang}
\end{center} 
\end{figure}

 
The antenna prototype was tested inside an Anechoic chamber with an Agilent Network Analyzer.  A real-time, over-the-air 2.4 GHz high frequency test signal was used to measure both the isolation and phase effects of the patch antenna.  We denote the measured isolation of the antenna in the passband by $|H^{RF}_{+}(f)|$, shown in Fig.~\ref{fig:chan_mag}, and the phase of the antenna by $\measuredangle H^{RF}_{+}(f)$, shown in Fig.~\ref{fig:chan_ang}.  The antenna is optimized for a carrier frequency of $f_c = 2.438$ GHz and measurements were made over a bandwidth $B_H = 20$ MHz centered at that frequency.  

Using those measurements, we can analytically express the self-interference channel as
\begin{equation}
H^{RF}_{+}(f) =
\left\{ \begin{array}{cl}
|H^{RF}_{+}(f)|e^{j\measuredangle H^{RF}_{+}(f)},  & |f - f_c| \leq \dfrac{B_H}{2}\\
0, & \textrm{elsewhere}
\end{array}\right.
\end{equation}
which is the one-sided FFT of the passband channel.  Using properties of the Fourier transform, we can write
\begin{equation}
H_{aa}(f) = \dfrac{1}{2}H^{RF}_{+}(f+f_c),
\label{eq:H_si_f}
\end{equation}
which is the FFT of the equivalent baseband channel centered at 0 Hz.  The time domain representation of the self-interference channel can finally be written as
\begin{equation}
h_{aa} = \mathcal{F}^{-1}\{H_{aa}(f) \},
\label{eq:h_si}
\end{equation}
after taking the IFFT. 

\section{Analog Baseband Cancellation}
\label{sec:cancel}
We now provide the details for the self-interference cancellation from the perspective of terminal $a$.  
\subsection{Channel Estimation}
In order to form an effective cancellation signal, the effects of the self-interference need to be estimated.  Each terminal will send training symbols while the other terminal remains silent.  Thus using (\ref{eq:r_bb}), the received signal at terminal $a$ during the training phase is
\begin{equation}
r_{a,tr} = \sqrt{P_{T_a}}h_{aa} x_{a,tr} + z_a,
\label{eq:r_train}
\end{equation}
and is achieved by terminal $a$ sending $N_{tr}$ training symbols with terminal $b$ silent.  Then using Least Squares channel estimation, terminal $a$ can form an estimate of the channel by
\begin{equation}
\widehat{h}_{aa} = \dfrac{r_{a,tr}x_{a,tr}^{-1}}{\sqrt{P_{T_a}}} = h_{aa} + \dfrac{z_a x_{a,tr}^{-1}}{\sqrt{P_{T_a}}},
\end{equation}
where the estimate $\widehat{h}_{aa}$ consists of the true channel corrupted by scaled additive noise.  The channel estimate can then be used to form the cancellation signal $\widehat{x}_a = -\sqrt{P_{T_a}} \: \widehat{h}_{aa} x_a$ which will attempt to cancel the self-interference signal $x_a$ during the data phase.  

\subsection{Self-Interference Cancellation}
Using the cancellation signal just derived with (\ref{eq:y_bb}), we can write
\begin{equation}
y_a = \sqrt{P_{T_b}}h_{ba}x_b + \sqrt{P_{T_a}}(h_{aa} -\widehat{h}_{aa}) x_{a} + z_a,
\label{eq:y_a}
\end{equation}
which is the received analog baseband signal at terminal $a$ after cancellation.  We define the unwanted residual self-interference signal at node $a$ as
\begin{equation}
y_{a,res} \triangleq \sqrt{P_{T_a}}(h_{aa} -\widehat{h}_{aa}) x_{a} + z_{a},
\label{eq:y_a_res}
\end{equation}
and notice that the power of the of the residual $E[|y_{a,res}|^2]$ increases proportionally with channel estimation error.  

We introduce two labels, PS and PS+B, to distinguish between the two different modes of self-interference cancellation available to the full-duplex terminals.  We use PS to denote when the terminals use only the patch antenna prototype for passive RF cancellation.  We then use PS+B to denote when the proposed analog baseband cancellation is used in combination with the patch antenna.  

\subsection{Results}
We now simulate the performance of the analog baseband cancellation at terminal $a$ using the scheme just described above.  We will use the \emph{Signal-to-Interference-Noise} (SINR) ratio as the main metric in order to quantify the strength of the desired signal over the combined self-interference and noise.  If we look at the SINR at terminal $a$
\begin{equation}
\Gamma^a = \dfrac{E[|\sqrt{P_{T_b}}h_{ba}x_b|^2]}{E[|y_{a,res}|^2]},
\label{eq:sinr_a}
\end{equation}
we define the strength of the desired signal as $P_{R_b} \triangleq E[|\sqrt{P_{T_b}}h_{ba}x_b|^2]$ in order parameterize it by a single value.   

\begin{figure}[htp]
\begin{center} 
  \includegraphics[width = 0.47\textwidth]{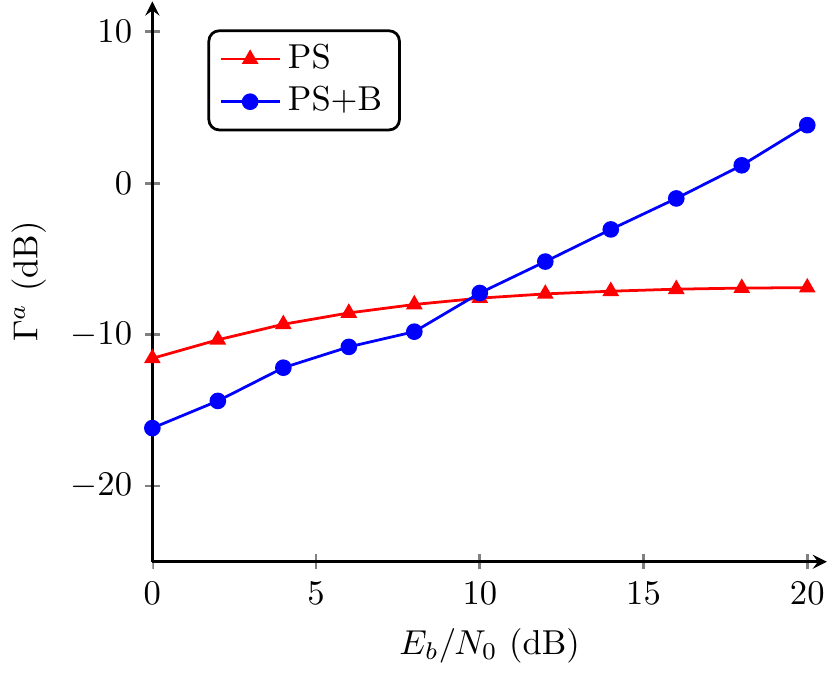}
  \vspace{-8pt}
  \caption[fig:rf_model]{The signal-to-interference-noise ratio ($\Gamma^a$) of the desired signal from terminal $b$ to the residual self-interference at terminal $a$.  Baseband cancellation PS+B improves the SINR as compared to the RF only PS scheme.} 
  \label{fig:SIR_add_B}
\end{center} 
\end{figure}

The performance of the cancellation scheme was simulated in Matlab using the experimental measurements of the patch antenna prototype for the self-interference channel.  Table~\ref{table:vars} shows the other relevant system parameters.  Terminal $a$ forms a channel estimate with $N_{tr}$ training symbols and then both terminals exchange $N_{bits}$ of bits with each other.  
\begin{table}[h]
\begin{center}  
\caption{Network Simulation Parameters} 
\centering 
\begin{tabular}{|c||c|} 
\hline   
\textbf{System Parameters} & \textbf{Value} \\   
\hline\hline 
PSK Modulation Order ($M$)& 4 \\
\hline
Number of Data Bits ($N_{bits}$) & 2000\\ 
\hline
Number of Training Symbols ($N_{tr}$)& $5$\\ 
\hline
Carrier Frequency ($f_c$)& 2.438 GHz  \\
\hline
Sampling Frequency ($F_s$)& 20 MHz  \\
\hline
Channel Bandwidth ($B_H$)& 20 MHz  \\
\hline
Signal Bandwidth ($BW$)& 10 MHz\\
\hline
Terminal $a$'s Transmit Power ($P_{T_a}$)& 0 dBm\\
\hline
Received Power from Terminal $b$ ($P_{R_b}$)& -60 dBm\\
\hline 
\end{tabular} 
\label{table:vars}  
\end{center}
\end{table} 
In Fig.~\ref{fig:SIR_add_B} the SINR is plotted versus $E_b/N_0$ to show the benefit of adding baseband cancellation to the RF passive cancellation provided by the patch antenna.

At approximately $E_b/N_0 = 10$ dB, the SINR for the RF PS scheme begins to saturate while the SINR for the baseband PS+B scheme continues to increase linearly with $E_b/N_0$.  The intersection point of the two curves is explained by the effect of noise on the baseband cancellation scheme.  In the presence of significant noise, $E_b/N_0 < 10$ in this scenario, the channel estimate will have high error and can actually cause more harm than good when forming the cancellation signal.  This affect can be observed how the PS scheme achieves higher SINR than the PS+B scheme.  For $E_b/N_0 > 10$, the affects of the additive noise lessen and the channel estimate can be used to create a beneficial cancellation signal.  We note that the value of $E_b/N_0$ is the same for both the transmitted data signals $x_a$ and $x_b$.

\begin{figure}[htp]
\begin{center} 
  \includegraphics[width = 0.47\textwidth]{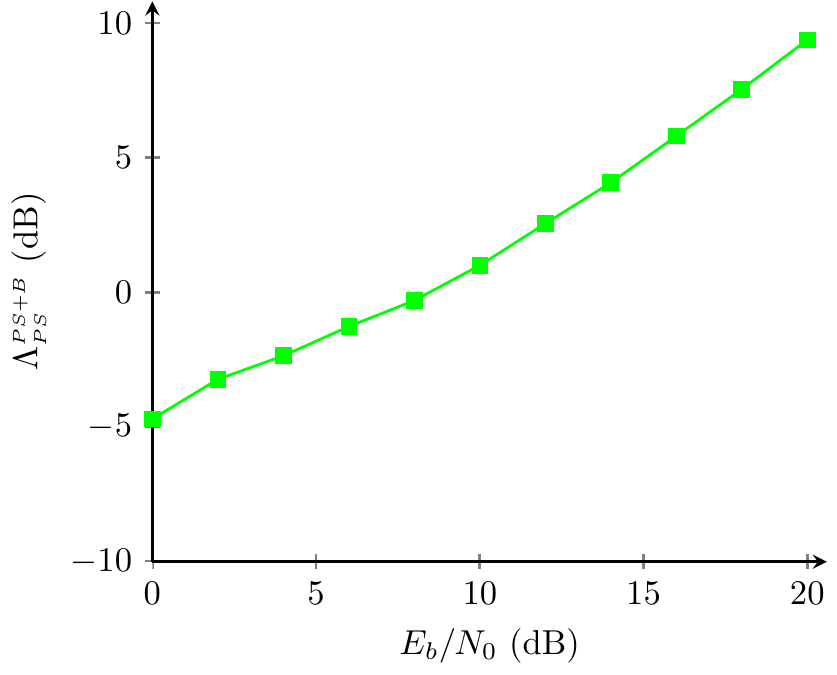}
  \vspace{-8pt}
  \caption[fig:chan_ang]{The relative SINR gain of the RF only cancellation scheme PS over the baseband cancellation scheme PS+B versus $E_b/N_0$.} 
  \label{fig:SIR_GAIN}
\end{center} 
\end{figure}

In order to quantify the tradeoff between the RF only cancellation of the PS scheme and the analog baseband cancelation scheme PS+B, we define the ratio 
\begin{equation}
\Lambda^{PS+B}_{PS} = \dfrac{\Gamma^a_{PS+B}}{\Gamma^a_{PS}},
\end{equation}
which calculates the relative SINR gain of the PS+B scheme over the PS scheme.  In Fig.~\ref{fig:SIR_GAIN}, we plot the SINR gain versus $E_b/N_0$.  We can immediately see that the SINR gain is linearly increasing proportional to the increasing strength of the desired signal when analog baseband cancellation is used in combination with the RF cancellation provided by the patch antenna.  
The zero-gain point is the same $E_b/N_0=10$ point discussed above. When the additive noise is too large, the RF only PS scheme outperforms the baseband scheme PS+B by 5 dB.  However, in low noise situations, the baseband scheme realizes gains up to 10 dB for the range of $E_b/N_0$ considered. 

\section{Full-Duplex Link Evaluation}
\label{sec:FDlink}
In the above section, we quantified the performance of the analog baseband cancellation scheme in terms of the signal strength of the desired signal.  We now quantify the performance of the point-to-point full-duplex link between terminals $a$ and $b$ with two different metrics.  The first performance metric is the classical Shannon information theoretic notion \cite{cover_thomas} of the achievable rate, $R^a= \log_2 (1+\Gamma^a)$, where the rate is measured in units of bits per second per Hertz (bps/Hz).   
\begin{figure}[htp]
\begin{center} 
  \includegraphics[width = 0.47\textwidth]{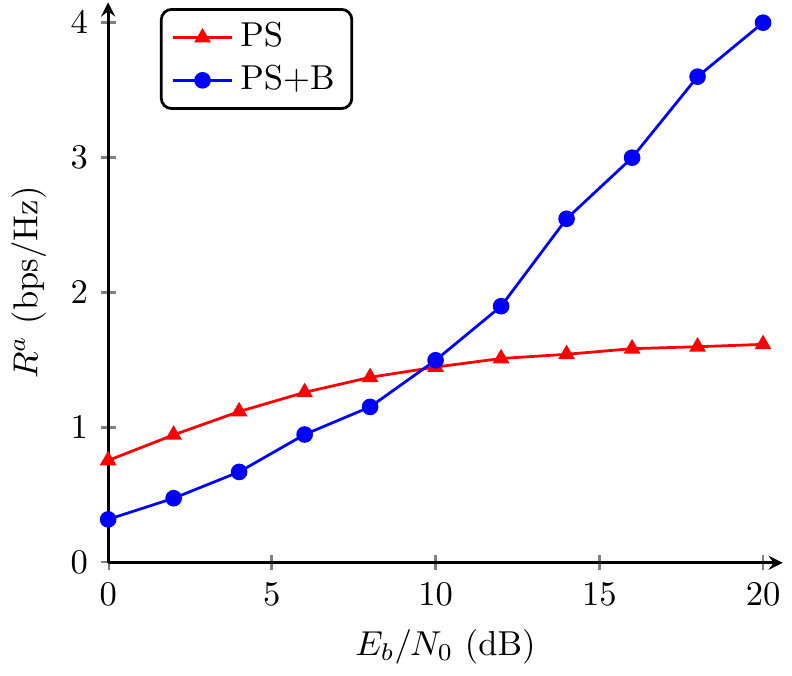} 
  \vspace{-10pt}
  \caption[fig:chan_ang]{The achievable rate ($R^a$) at terminal $a$ versus $E_b/N_0$.  The baseband PS+B scheme achieves almost 2.5 bps/Hz higher rate as compared to the RF PS scheme.} 
  \label{fig:rate} 
\end{center} 
\end{figure}    
In Fig.~\ref{fig:rate}, we plot the achievable rate at terminal $a$ as a function of $E_b/N_0$.  Because the rate is a function of the SINR value $\Gamma^a$ defined and evaluated above, we see similar trends in the rate of the full-duplex link as were observed for the SINR.  The rate of the RF cancellation PS scheme saturates at about 1.5 bps/Hz.  The achievable rate of the baseband PS+B scheme is linearly increasing with $E_b/N_0$ and can achieve up to 4 bps/Hz.  

\begin{figure}[htp]
\begin{center} 
  \includegraphics[width = 0.47\textwidth]{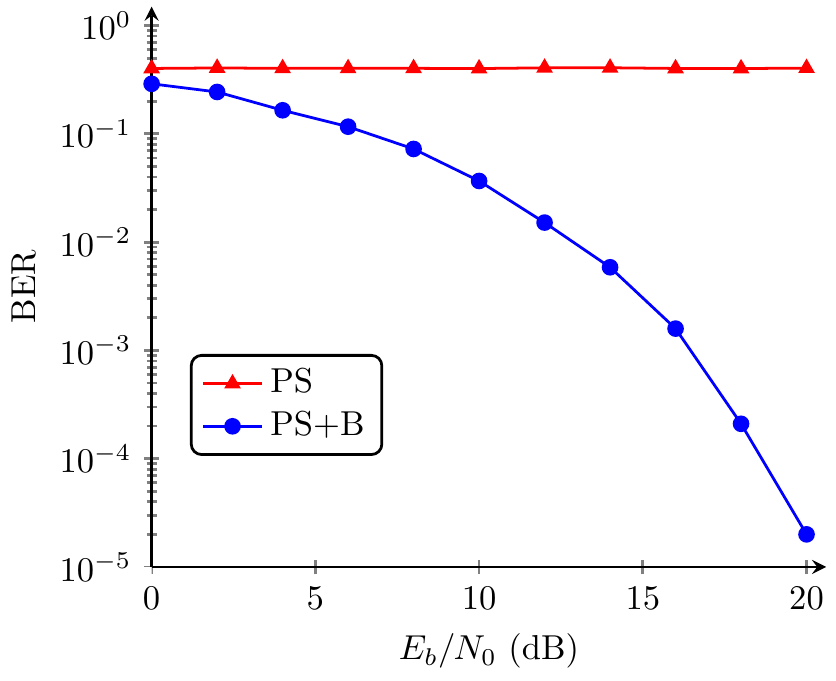} 
  \caption[fig:chan_ang]{The bit error rate (BER) at terminal $a$ versus $E_b/N_0$.  The analog baseband cancellation scheme PS+B significantly reduces the error rate as compared to the RF only cancellation scheme PS.} 
  \label{fig:BER} 
\end{center} 
\end{figure}

The second metric we consider is the bit error rate (BER) of the bits transmitted by terminal $b$ and received by terminal $a$. Fig.~\ref{fig:BER} plots the BER with respect to $E_b/N_0$.  It is clearly noticeable how the use of baseband cancellation improves the link quality.  For $E_b/N_0 = 10$ dB, we see a factor of 10 improvement in the BER and at $E_b/N_0 = 20$ dB, up to $10^4$ improvement is observed. 

\section{Conclusion}
\label{sec:conclusion}
This paper proposes and evaluates an analog baseband self-interference cancellation scheme.  Real-time, over-the-air measurements of a four-layer RF patch antenna prototype were used to characterize the RF self-interference channel.  The channel model combined with a practical transceiver model enables us to derive an analytical baseband signal model incorporating the RF self-interference effects.  Least squares channel estimation in the analog baseband stage of the receiver is used to estimate the self-interference channel and generate a cancellation signal just prior to the analog-to-digital converter.  The performance of the cancellation scheme was quantified through the SINR ratio of the desired signal with respect to the residual self-interference signal.  The analog baseband cancellation scheme PS+B achieves up to 10 dB higher SINR than the RF only cancellation scheme PS.  We then evaluate the performance of a point-to-point full-duple link with the achievable rate and BER used as metrics.  The baseband PS+B scheme is able to achieve up to 2.5 bps/Hz improvement in achievable rate as compared to the PS scheme.  A $10^1-10^4$ reduction in BER was achieved by adding analog baseband cancellation to the RF only cancellation scheme. 

These initial results provide motivation for adding analog baseband self-interference cancellation to current systems that only employ RF self-interference cancellation.  Our proposed baseband cancellation scheme is agnostic to the specific RF self-interference channel model and can be utilized with various other channel models.  In our own extensions of this work, we consider alternate channel models in \cite{bkaufman_j3} and provide more in depth analysis.  

\vspace{-10pt}
\section{Acknoldgements}
The authors would like to thank Dr. Aydin Babakhani and graduate student researchers Tulong Yang and Peiyu Chen.  The patch antenna prototype was developed in their lab, Rice Integrated Systems and Circuits (RISC), at Rice University.

\bibliographystyle{IEEEtran}   
\bibliography{/Users/bkaufman/Dropbox/Brett/Brett_Papers/Bibs/full_duplex_brett.bib,bkaufman_temp.bib,/Users/bkaufman/Dropbox/Brett/Brett_Papers/Bibs/Brett_Textbooks.bib}   

\begin{thebibliography}{10}
\providecommand{\url}[1]{#1}
\csname url@samestyle\endcsname
\providecommand{\newblock}{\relax}
\providecommand{\bibinfo}[2]{#2}
\providecommand{\BIBentrySTDinterwordspacing}{\spaceskip=0pt\relax}
\providecommand{\BIBentryALTinterwordstretchfactor}{4}
\providecommand{\BIBentryALTinterwordspacing}{\spaceskip=\fontdimen2\font plus
\BIBentryALTinterwordstretchfactor\fontdimen3\font minus
  \fontdimen4\font\relax}
\providecommand{\BIBforeignlanguage}[2]{{%
\expandafter\ifx\csname l@#1\endcsname\relax
\typeout{** WARNING: IEEEtran.bst: No hyphenation pattern has been}%
\typeout{** loaded for the language `#1'. Using the pattern for}%
\typeout{** the default language instead.}%
\else
\language=\csname l@#1\endcsname
\fi
#2}}
\providecommand{\BIBdecl}{\relax}
\BIBdecl

\bibitem{FD_Radar}
F.~O'Hara and G.~Moore, ``A high performance {CW} receiver using feed thru
  nulling,'' \emph{Microwave Journal}, Sep. 1963.

\bibitem{FD_Repeater}
R.~Isberg and W.~Lee, ``Performance tests of a low power cellular enhancer in a
  parking garage,'' in \emph{IEEE Vehicular Technology Conference (VTC)}, 1989.

\bibitem{2010_Melissa_Asilomar}
M.~Duarte and A.~Sabharwal, ``Full-duplex wireless communications using
  off-the-shelf radios: Feasibility and first results,'' in \emph{Asilomar
  Conf. on Signals, Systems, and Comp.}, Nov. 2010.

\bibitem{2010_Stanford_Mobicom}
J.~Choi, M.~Jain, K.~Srinivasan, P.~Levis, and S.~Katti, ``Achieving single
  channel, full duplex wireless communication,'' in \emph{MOBICOM}, Sep. 2010.

\bibitem{2013_Evan_TWC_passive}
\BIBentryALTinterwordspacing
E.~Everett, A.~Sahai, and A.~Sabharwal, ``Passive self-interference suppression
  for full-duplex infrastructure nodes,'' Jan. 2013, submitted to \emph{IEEE
  Transactions on Wireless Communication}. [Online]. Available:
  \url{http://arxiv.org/abs/1302.2185}
\BIBentrySTDinterwordspacing

\bibitem{2012_MIDU}
E.~Aryafar, M.~Khojastepour, K.~Sundaresan, S.~Rangarajan, and M.~Chiang,
  ``{MIDU}: enabling {MIMO} full duplex,'' in \emph{MOBICOM}, 2012.

\bibitem{2010_Patch_Sweden}
K.~Haneda, E.~Kahra, S.~Wyne, C.~Icheln, and P.~Vainikainen, ``Measurement of
  loop-back interference channels for outdoor-to-indoor full-duplex radio
  relays,'' in \emph{European Conf. on Antennas and Propagation (EuCAP)}, Apr.
  2010.

\bibitem{2012_Melissa_TWC}
M.~Duarte, C.~Dick, and A.~Sabharwal, ``Experiment-driven characterization of
  full-duplex wireless systems,'' \emph{IEEE Trans. on Wireless Commun.},
  vol.~11, no.~12, Dec. 2012.

\bibitem{2013_Stanford_Sigcomm}
D.~Bharadia, E.~McMilin, and S.~Katti, ``Full duplex radios,'' in \emph{ACM
  SIGCOMM}, Aug. 2013.

\bibitem{RISC_WEB}
\BIBentryALTinterwordspacing
Rice Integrated Systems and Circuits Lab (RISC). [Online]. Available:
  \url{http://www.ece.rice.edu/~ab28/index.html}
\BIBentrySTDinterwordspacing

\bibitem{2011_patch_antenna}
J.~Lu, Z.~Kuai, X.~Zhu, and N.~Zhang, ``A high-isolation dual-poloarization
  microstrip patch antenna with quasi-cross-shaped coupling slot,'' \emph{IEEE
  Trans. Antennas Propag.}, vol.~59, no.~7, Jul. 2011.

\bibitem{cover_thomas}
T.~M. Cover and J.~A. Thomas, \emph{Elements of Information Theory}.\hskip 1em
  plus 0.5em minus 0.4em\relax John Wiley {\&} Sons, 1991.

\bibitem{bkaufman_j3}
B.~Kaufman, J.~Lilleberg, and B.~Aazhang, ``Analog baseband cancellation for
  full-duplex: An experiment driven analysis,'' \emph{submitted to IEEE Journal
  on Selected Areas of Communications (JSAC) Special Issue on Full-Duplex
  Communications and Networks}, October 2013.

\end{thebibliography}

\end{document}